\documentstyle[12pt]{article}
\begin{document}
\def\thefootnote{\fnsymbol{footnote}}
\begin{flushright}
KANAZAWA-06-06  \\
June, 2006
\end{flushright}
\vspace*{2cm}
\begin{center}
{\LARGE\bf SUSY breaking based on Abelian gaugino kinetic term mixings}\\
\vspace{1 cm}
{\Large Daijiro Suematsu}
\footnote[2]{e-mail:~suematsu@hep.s.kanazawa-u.ac.jp}
\vspace {1cm}\\
{\it Institute for Theoretical Physics, Kanazawa University,\\
        Kanazawa 920-1192, Japan}\\
\end{center}
\vspace{1cm}
{\Large\bf Abstract}\\
We present a SUSY breaking scenario based on Abelian 
gaugino kinetic term mixings between hidden and observable 
sectors. If an extra U(1) gaugino in the observable sector obtains
a large mass through this mixing effect based on SUSY breaking 
in the hidden sector, soft SUSY breaking parameters in the 
MSSM may be affected by radiative effects due to this gaugino mass. 
New phenomenological aspects are discussed in such a SUSY 
breaking scenario. 

\newpage
\setcounter{footnote}{0}
\def\thefootnote{\arabic{footnote}}
Supersymmetry (SUSY) is considered to be the most promising solution for 
the gauge hierarchy problem of the standard model (SM) \cite{rev}.  
Since phenomenological features of SUSY models are 
determined by soft SUSY breaking parameters,
it is crucial to clarify the nature of SUSY breaking mechanisms.
Since a favorable SUSY breaking scale in a hidden sector 
depends on what kind of mediation scenarios of SUSY breaking are
supposed, features of the SUSY breaking in the observable sector 
is usually fixed by a
dominant contribution due to a certain mediation among them. 
If different mediation mechanisms can compete to induce 
the SUSY breaking in the observable sector, there may appear
novel feature in SUSY breaking parameters.
In this paper we discuss a possibility that the coexistence of different
mediations of the SUSY breaking may break the universality of 
gaugino masses and  bring new
phenomenological features to the models.

A few examples which can realize non-universal gaugino masses have been 
proposed,\footnote{The gaugino masses are known to
be non-universal in some types of models, for example, 
in the multi-moduli SUSY breaking \cite{multim}, intersecting 
D-brane models \cite{dbrane} and a certain type of gauge mediation models
\cite{s}. } and its phenomenological consequences have been examined
from some view points \cite{multim}-\cite{cdm}. 
Here we propose a new mediation mechanism of the SUSY 
breaking in the hidden sector to the observable sector, which can 
make an Abelian gaugino mass largely different from others in the
observable sector. In such a case other SUSY breaking parameters 
may be also affected through radiative effects of this large 
Abelian gaugino mass. In particular, corrections to the Higgs and stop
masses seem to be phenomenologically interesting since it may help to
soften the little hierarchy problem in the MSSM.

In this paper we use the Abelian gauge kinetic term mixing for the
mediation of the SUSY breaking.
It is known that the kinetic term mixing can generally occur 
among the Abelian gauge fields in the models with multi U(1)s 
\cite{mixing,oneloop,mixing1}.
We assume that such mixing exists between two Abelian 
gauge fields, one of which belongs to the hidden sector and the other 
belongs to the observable sector.
In that case we show that there can be an additional contribution to 
the corresponding Abelian gaugino mass in the observable sector, 
if certain assumptions for the superpotential and the 
SUSY breaking in the hidden sector are satisfied.   
This additional contribution can make the Abelian
gaugino mass different from others in the observable sector.
Moreover, it may bring radiatively dominant contributions 
for certain SUSY breaking parameters depending on the SUSY 
breaking scale in the hidden sector.

The following parts are organized as follows.
First, we explain the SUSY breaking mediation due to the gauge kinetic
term mixing. We discuss how the gaugino mass of an additional U(1)
factor in the observable sector can be heavier than other gauginos. 
After that, we estimate the
corrections to the Higgs and stop masses due to this gaugino mass.
We discuss these corrections may help to soften the little hierarchy
problem in the MSSM. We give numerical results of such
analyses for the extra U(1) models derived from $E_6$, as an example.
 
For simplicity, we consider a SUSY U(1)$_a\times$U(1)$_b$ model
where U(1)$_a$ and U(1)$_b$ belong to the hidden and observable 
sectors, respectively. In the later discussion U(1)$_b$ is identified
with an additional U(1)$_x$ to the MSSM.
We suppose that $\hat W_{a,b}^\alpha$ is a chiral superfield with a spinor
index $\alpha$, which contains the field strength of U(1)$_{a,b}$. 
Since $\hat W_{a,b}^\alpha$ is gauge invariant by itself, the gauge invariant 
kinetic terms can be expressed as
\begin{equation}
{\cal L}_{\rm kin}=\int d^2\theta\left({1\over 32} 
\hat W_a^\alpha\hat W_{a\alpha} 
+{1\over 32}\hat W_b^\alpha\hat W_{b\alpha} 
+ {\sin\chi\over 16}\hat W_a^\alpha\hat W_{b\alpha} \right).
\label{kinetmix}
\end{equation}
A mixing term is generally allowed 
at least from a viewpoint of the symmetry.
Although some origins such as the string one-loop effect may be
considered for this mixing \cite{oneloop}, 
we only treat $\sin\chi$ in eq.~(\ref{kinetmix}) 
as a free parameter.

This mixing can be resolved by practicing a transformation
\cite{mixing}
\begin{equation}
\left(\begin{array}{c}\hat W_a^\alpha \\ \hat W_b^\alpha \end{array}\right)
=\left(\begin{array}{cc} 1 & -\tan\chi \\ 0 & 1/\cos\chi \\\end{array}
\right)\left(\begin{array}{c}\hat W_h^\alpha \\
\hat W_x^\alpha \end{array}\right).
\label{base}
\end{equation}
If we use a new basis $(\hat W_h^\alpha, \hat W_x^\alpha)$, 
a covariant derivative
in the observable sector can be written as
\begin{equation}
D^\mu=\partial^\mu+i\left(-g_aQ_a\tan\chi+{g_bQ_b\over\cos\chi}
\right)A_x^\mu.
\end{equation}
This shows that the gauge field $A_x^\mu$ in the observable sector can
interact with the hidden sector fields
which have a nonzero charge $Q_a$.
However, since the fields in the hidden sector are 
generally considered to be heavy
enough and $\sin\chi$ is expected to be small, we can safely expect 
that there is no phenomenological contradiction at the present stage. 

We consider that the Abelian gauginos in both sectors obtain masses 
through the SUSY breaking in the hidden sector such as 
\begin{equation}
{\cal L}_{\rm gaugino}^m=M_a\tilde\lambda_a\tilde\lambda_a
+M_b\tilde\lambda_b\tilde\lambda_b,
\label{mgauge}
\end{equation}
where the mass $M_b$ of the gaugino in the observable sector may 
be supposed as the ordinary universal mass $m_{1/2}$. 
If we can assume that $M_a \gg M_b$ is satisfied,
these mass terms are rewritten by using the new basis (\ref{base}) as follows,
\begin{equation}
\tilde{\cal L}_{\rm gaugino}^{m} 
=M_a\tilde\lambda_h\tilde\lambda_h+(M_b
+M_a\sin^2\chi)\tilde\lambda_x\tilde\lambda_x,
\label{modmass}
\end{equation}
where we also use $\sin\chi\ll 1$ in this derivation.
This suggests that the mass of the Abelian gaugino in the observable sector 
can have an additional 
contribution due to the gauge kinetic term mixing with the
gaugino in the hidden sector.
This new contribution can be a dominant one when 
the SUSY breaking in the hidden sector satisfies 
\begin{equation}
M_a\sin^2\chi>M_b.
\label{mcond}
\end{equation} 
In this case the universality of the gaugino masses in the observable
sector can be violated in the Abelian part.

Next we present an example of the SUSY breaking scenario 
which can satisfy the condition (\ref{mcond}) in the framework of the 
gravity mediation SUSY breaking. 
We consider the hidden sector which contains chiral superfields 
$\hat\Phi_{1,2}$ charged under U(1)$_a$. It is also supposed to contain 
various neutral chiral superfields like moduli, which are represented by
$\hat{\cal M}$ together. They are defined as dimensionless fields. 
Matter superfields in the observable sector are denoted by $\hat\Psi_I$.
Both K\"ahler potential and superpotential relevant to the
present argument are supposed to be written as\footnote{For simplicity,
we assume minimal kinetic terms for the matter fields. A hat is put on for
the superfield and the scalar component is represented by removing the
hat from it.}
\begin{eqnarray}
&&{\cal K}=\kappa^{-2}\hat K(\hat{\cal M})+\hat\Phi_1^\ast\hat\Phi_1
+\hat\Phi_2^\ast\hat\Phi_2+\hat\Psi_I^\ast\hat\Psi_I+\cdots, \nonumber\\
&&{\cal W}=\hat W_0(\hat M)+\hat W_1(\hat{\cal M})\hat\Phi_1\hat\Phi_2
+{1\over 3}\hat Y_{IJK}\hat\Psi_I\hat\Psi_J\hat\Psi_K+\cdots,
\end{eqnarray} 
where $\kappa^{-1}$ is the reduced Planck mass and 
$Q_a(\hat\Phi_1)+Q_a(\hat\Phi_2)=0$ is assumed. 
As a source relevant to the SUSY breaking in the hidden 
sector, we adopt a usual assumption in case of the gravity mediation 
SUSY breaking. That is, the SUSY breaking effects are
assumed to be parameterized by $F$-terms $F_{\cal M}$ of certain moduli 
${\cal M}$ \cite{sgrasusy}. In that case the
gravitino mass $m_{3/2}$ can be defined by $m_{3/2}\equiv\kappa^2e^{K/2}W_0$.
Since vacuum energy is expressed  by using these as
$V_0=\kappa^{-2}(F^{\cal M}F^{\bar{\cal M}}
\partial_{\cal M}\partial_{\bar{\cal M}}K -3m_{3/2}^2)$,
$F^{\cal M}$ is supposed to be $O(m_{3/2})$ as long as $V_0$ 
is assumed to vanish.

Applying this assumption to the scalar potential formula in the
supergravity, we can obtain well known soft supersymmetry breaking 
parameters for the scalar masses $m_{\tilde f}$ and three scalar couplings
$A_{\tilde f_1\tilde f_2\tilde f_3}$ in the observable sector as 
\cite{sgrasusy}
\begin{equation}
m^2_{\tilde f}= m_{3/2}^2, \qquad 
A_{\tilde f_1\tilde f_2\tilde f_3}=\sqrt{3}m_{3/2}. 
\label{softb1}
\end{equation}
The masses of the gauginos are generated as \cite{sgrag}
\begin{equation}
m_{1/2}={1\over 2Re[f_A({\cal M})]}
F^{\cal M}\partial_{\cal M} f_A({\cal M}),
\label{softb2}
\end{equation}
where $f_A({\cal M})$ is a gauge kinetic function for a gauge 
factor group $G_A$. If $f_A({\cal M})$ takes the same form 
for each factor group,
the universal gaugino masses are generated as
$m_{1/2}=O(m_{3/2})$. This is the ordinary scenario.
In the present case, the gaugino mass $M_b$ in eq.~(\ref{mgauge}) 
is expected to be induced from this gravity mediation and take 
the universal value $m_{1/2}$.

On the other hand, the gaugino mass $M_a$ in the hidden sector 
is generated through the mediation of charged chiral 
superfields $\hat\Phi_{1,2}$ due to 
the second term in ${\cal W}$ as in the gauge mediation SUSY 
breaking scenario \cite{gmed}. 
Since it can be generated by one-loop diagrams which have
component fields of $\hat\Phi_{1,2}$ in internal lines, it is
approximately expressed as
\begin{equation}
M_a={g_a^2\over 16\pi^2}\Lambda,
\label{mgauge1}
\end{equation}
where $\Lambda=\langle F_1\rangle/\langle S_1\rangle $ and
we define that $S_1$ and $F_1$ are the scalar and auxiliary
components of $\hat W_1$, respectively.
Since we are considering the gravity mediation SUSY breaking,
the SUSY breaking scale in the hidden sector should be large 
to induce a suitable breaking in the observable sector.
Since $\Lambda=O((\kappa^{-1}m_{3/2})^{1/2})$ is required and then
$M_a$ can be much larger than the gravity mediation
contribution $M_b=O(m_{1/2})$, the additional contribution $M_a\sin^2\chi$ 
to the Abelian gaugino mass in eq.~(\ref{modmass}) 
can break the gaugino mass universality largely in the observable sector. 
In fact, if $\sin\chi$ takes a suitable value 
such as $\chi=O(10^{-1})$,\footnote{The 
string one-loop effect may bring this order of mixing as 
discussed in \cite{oneloop}.}  
we can expect that $M_a\sin^2\chi > M_b$ is realized 
and the Abelian gaugino mass characterized by $M_a\sin^2\chi$
can take a much larger value than other universal ones.\footnote{If 
the absolute values of $M_b$ and
$M_a\sin^2\chi$ could be the same order, two contribution
might substantially cancel each other to realize much smaller 
value than $m_{1/2}$. Although this may be an interesting possibility,
we do not consider it here.}

Now we discuss an interesting consequence of this kind
of scenario for the little hierarchy problem in the MSSM. 
The large mass of the Abelian gaugino in the observable sector 
can induce additional corrections
to other soft SUSY breaking parameters through the
renormalization group equations (RGEs).
In order to fix the features of this SUSY breaking scenario,
we need to examine this radiative effects and compare them with the
SUSY breaking effects due to the gravity mediation.
Radiative collections to other soft SUSY breaking parameters 
induced by the mass $M_x$ of the U(1)$_x$ gaugino 
can be estimated by solving the RGEs. 
The gaugino mass $M_x$ runs as
\begin{equation}
M_x(M_w)={\alpha_x(M_w)\over \alpha_x(\Lambda)}M_x(\Lambda), 
\label{gmass} 
\end{equation}
where $\alpha_x=g_x^2/4\pi$ and 
$\alpha_x(M_w)/\alpha_x(\Lambda)=1-(b_x\alpha_x(M_w)/ 2\pi)
\ln(\Lambda/M_w)$. The scale $\Lambda$ is introduced in
eq.~(\ref{mgauge1}) and $M_w$ is the weak scale. 
Since the large Abelian gaugino mass $M_x$ gives dominant contributions
in the RGEs for the soft SUSY breaking parameters, 
we may approximately estimate their evolution by taking account of 
its effect alone.   
Using one-loop RGEs, the soft masses of the scalar components 
$\tilde f$ and the $A$ parameters for the couplings of three scalars
 $\tilde f_1\tilde f_2\tilde f_3$ are represented as
\begin{eqnarray}
m^2_{\tilde f}(M_w)&\simeq& \bar m^2_{\tilde f}(M_w)
-{Q_{\tilde f}^2 \over 2b_x}
\left({\alpha_x^2(M_w)\over\alpha_x^2(\Lambda)}-1\right)M_x^2(\Lambda), 
\nonumber\\
A_{\tilde f_1\tilde f_2\tilde f_3}(M_w)&\simeq&
\bar A_{\tilde f_1\tilde f_2\tilde f_3}(M_w)
-{(\sum_{i=1}^3Q_{\tilde f_i}^2)\over 2b_x}
\left({\alpha_x(M_w)\over\alpha_x(\Lambda)}-1\right)
M_x(\Lambda),
\label{softmass}
\end{eqnarray}
where $Q_\psi$ stands for the U(1)$_x$ charge of the field $\psi$.
The MSSM effects based on the gravity mediation are summarized 
by $\bar m_{\tilde f}^2$ and 
$\bar A_{\tilde f_1\tilde f_2\tilde f_3}$.
Since the correction induced by $M_x$ to the gaugino mass $M_A$ of 
the factor group $G_A$ appears as two-loop effects 
\cite{twoloop}, we can safely estimate $M_A$ by taking account of the
gravity mediation effect only and using the
one-loop RGE formula given in eq.~(\ref{gmass}). 
If we note that $Q_\psi=O(1)$ and $\Lambda$ is an intermediate scale,
the first and second terms of $m_{\tilde f}^2(M_w)$ in 
eq.~(\ref{softmass}) are expected to take similar
order values as long as $M_x$ is $O(1)$~TeV. Thus, additional
corrections to the soft scalar masses may work to improve the degeneracy
among the soft scalar masses and then suppress flavor changing neutral currents
(FCNC) as long as U(1)$_x$ is generation blind. 
Corrections due to $M_x$ for the $A$ parameter is expected to be smaller than 
the gravity induced one. 

This additional correction to the scalar masses may improve the situation
for the radiative symmetry breaking in the MSSM. 
The potential minimum conditions in the MSSM can be written as
\begin{equation}
\sin 2\beta={2B\mu \over m_1^2+ m_2^2 +2\mu^2}, \qquad
m_Z^2={2m_1^2-2m_2^2\tan^2\beta\over \tan^1\beta-1}-2\mu^2.
\label{rad}
\end{equation}
Since the Higgs mass difference $\delta\equiv m_1^2-m_2^2$ at the top mass
scale $m_t$ can be approximately written by
taking account of the stop loop effect and the formulas in 
eq.~(\ref{softmass}) as
\begin{equation}  
\delta\simeq{3h_t^2\over 4\pi^2}
m_{\tilde t}^2\ln\left({m_{\tilde t}\over m_t}\right)
-{Q_1^2-Q_2^2 \over 2b_x}\left({\alpha_x^2(m_t)\over\alpha_x^2(\Lambda)}
-1\right)M_x^2(\Lambda),
\label{corr}
\end{equation}
the conditions (\ref{rad}) can be summarized as
\begin{equation}
m_Z^2=\left({\delta\over\tan^2\beta-1}-{B\mu\over\tan\beta}\right)
(\tan^2\beta+1).
\label{rad1}
\end{equation}
Although the present mass bounds for the lightest neutral Higgs scalar 
require a large stop mass $m_{\tilde t}$, the large stop mass imposes 
the tuning among the SUSY breaking parameters so as to satisfy
eq.~(\ref{rad1}). As a result, we have a so-called little hierarchy problem.
However, in the present case the situation may be changed.
The additional correction due to $M_x$ increases the stop mass 
through eq.~(\ref{softmass}). On the other hand, eq.~(\ref{corr}) shows
that it could reduce a value of $\delta$ in case of $Q_1^2<Q_2^2$ and
then it might relax the tuning required to satisfy eq.~(\ref{rad1}). 

We examine this aspect numerically in interesting examples which 
have U(1)$_x$ as an additional Abelian gauge symmetry to the MSSM. 
As such a U(1)$_x$ we adopt an Abelian symmetry derived from $E_6$  
and fix the model in the following way. 
U(1)$_x$ is identified with a linear 
combination of two additional U(1)'s to the MSSM as
\begin{equation}
Q_x=Q_\psi\cos\theta-Q_\chi\sin\theta,
\end{equation}
where $Q_\psi$ and $Q_\chi$ are the charges of U(1)$_\psi$ and
U(1)$_\chi$ in 
$$
E_6\supset SO(10)\times U(1)_\psi \supset 
SU(5)\times U(1)_\psi\times U(1)_\chi.
$$ 
Following eqs.~(\ref{softb1}) and (\ref{softb2}), the gravity mediation SUSY
breaking parameters are fixed as the universal ones 
$m_{\tilde f}=A_{\tilde f_1\tilde f_2\tilde f_3}/\sqrt{3}=m_{1/2}=m_{3/2}$.
The gaugino of U(1)$_x$ is assumed to obtain the
additional large mass $M_x$ through the kinetic term mixings with the 
hidden sector field. We assume the MSSM contents as chiral matter 
fields and impose the GUT normalization $g_x=\sqrt{5\over 3}g_1$. 
In this type of U(1)$_x$ extra matter fields from the {\bf 27}'s of
$E_6$ are required to cancel the anomaly. However, since their effects are
expected to be secondary, we do not take them into account here.

\input epsf 
\begin{figure}[tb]
\begin{center}
\epsfxsize=7.5cm
\leavevmode
\epsfbox{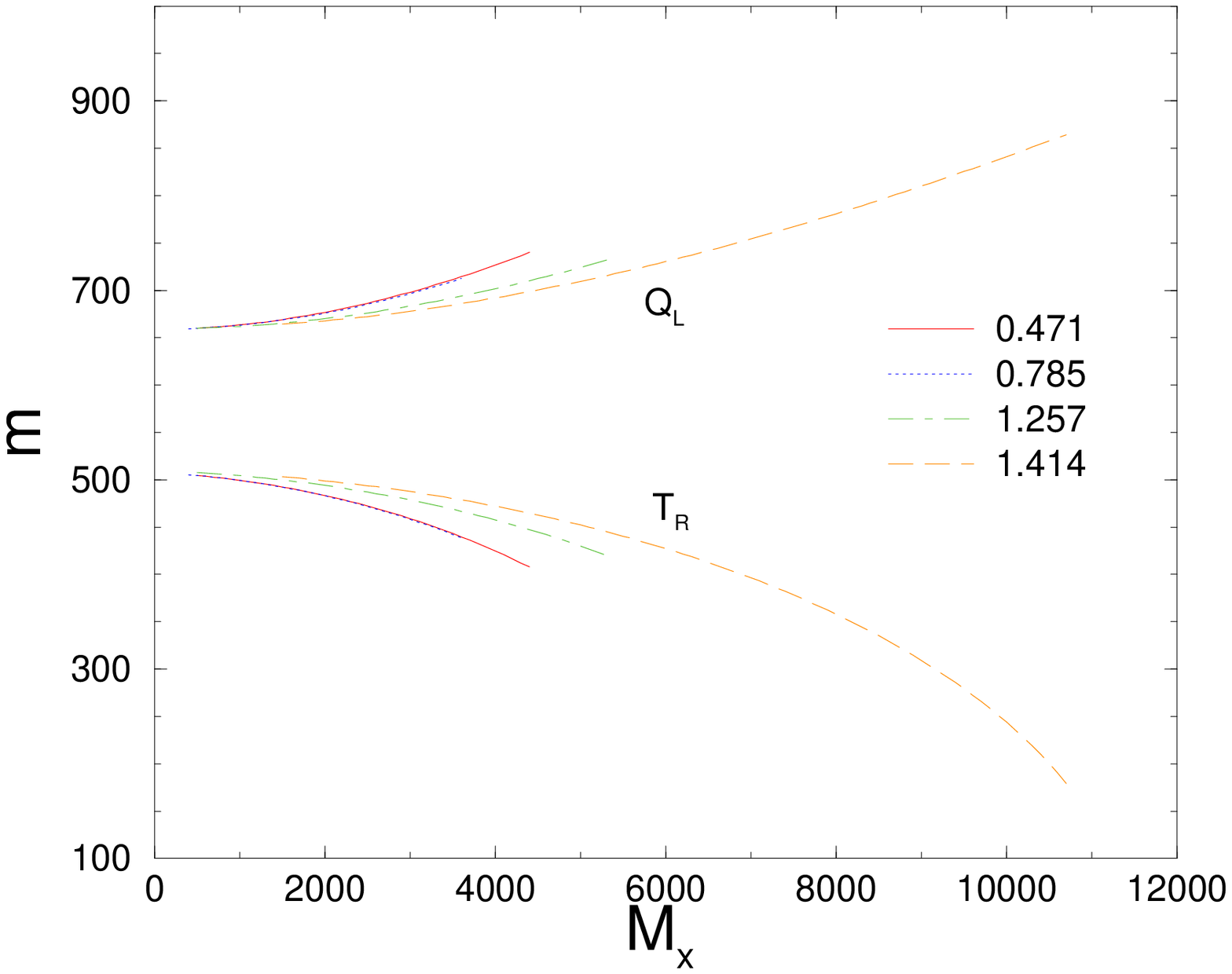}
\hspace*{3mm}
\epsfxsize=7.5cm
\leavevmode
\epsfbox{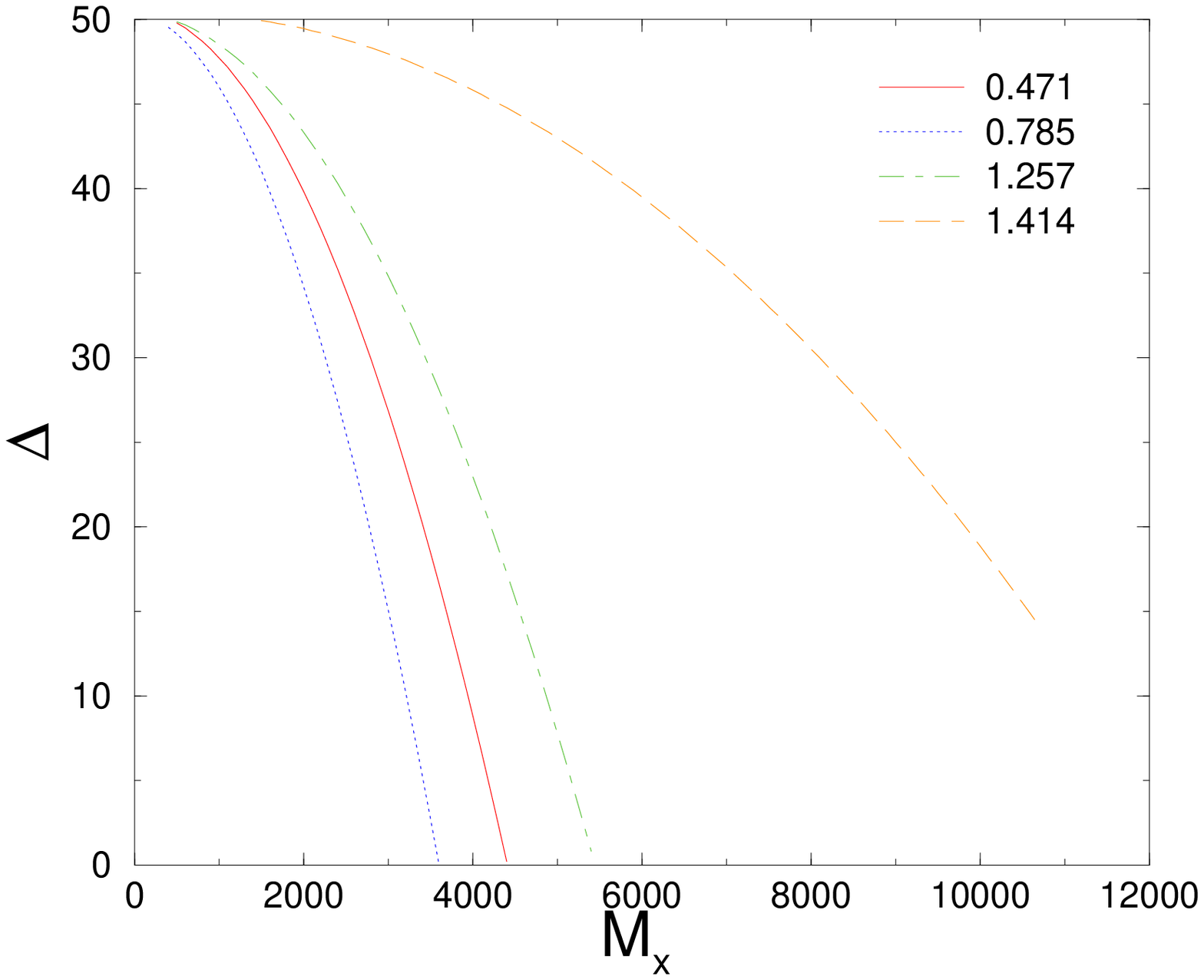}
\end{center}
\vspace*{-3mm}
{\footnotesize Fig.~1~~ One-loop corrections induced by the $M_x$
 effects for several values of $\theta$. 
In the left panel stop soft masses $m_{\tilde Q_L}$ 
and $m_{\tilde T_R}$ are plotted. In the right panel $\Delta$ is
 plotted. GeV is used as the mass unit.}
\end{figure}

We study the behavior of the SUSY breaking parameters of this model 
by using one-loop RGEs for $m_{3/2}=300$~GeV and $\Lambda=10^{11}$~GeV.
By varying the value of $M_x$ at the scale $\Lambda$ for various values of 
$\theta~({-\pi/2<\theta<\pi/2})$,
we calculate the stop masses and $\Delta\equiv\delta/m_Z^2$.  $\Delta$ is 
considered as a measure for the required fine tuning in eq.~(\ref{rad1}).   
Results in case of $\tan\beta=2$ are 
plotted for the several values of $\theta$ 
in Fig.~1.\footnote{There is U(1)$_x$ 
$D$-term contribution to the soft scalar masses such as 
$(m_{\tilde f}^{D_x})^2\simeq g_x^2Q_fQ_S\langle S\rangle^2$, 
where $\langle S\rangle$ determines a U(1)$_x$ breaking
scale. Since we fix $\langle S\rangle$ as $\langle S\rangle=1.5$~TeV and
then $M_x > g_x\langle S\rangle$ is satisfied in the almost all regions
of the large $M_x$, this contribution is neglected 
in this calculation. }
In the left panel we plot the stop masses $m_{\tilde Q_L}$ and 
$m_{\tilde T_R}$. The same input parameters give 
$m_{\tilde Q_L}\simeq 716$~GeV and $m_{\tilde T_R}\simeq 583$~GeV in
case of the MSSM with $\tan\beta=7$.
This panel shows that the averaged stop mass increases for the larger $M_x$
through the RGE effect as expected.
In the right panel we show the behavior of $\Delta$. 
For the same input parameters we find $\Delta\simeq 49$ in the MSSM.  
From this figure we find that the larger values of $M_x$ can make the value of
$\Delta$ smaller. This can be explained by the second term in
eq.~(\ref{corr}), which may cancel the contribution of the first term even
in the case of large stop mass. In fact,
we find that this can happen for suitable values of $\theta$ for which 
$Q_1^2<Q_2^2$ is satisfied as mentioned before. 
Even if the correction due to $M_x$ makes the stop mass increase, 
the present results show that $\Delta$ can be smaller satisfying 
the neutral Higgs mass bound as long as $Q_1^2<Q_2^2$ is satisfied.
The situation for the parameter tuning in eq.~(\ref{rad1}) seems to be
improved for rather wide ranges of the value of $M_x$. 

In the extra U(1) model studied here, it is also useful to note 
that the constraint 
from the neutral Higgs mass bounds can be relaxed by additional
contributions in comparison with the MSSM \cite{extra}. 
The constraint can be satisfied even for the small values of $\tan\beta$
such as 2, which is used in the above calculations.
This point may also be considered a favorable feature of this scenario 
for the fine tuning problem. In addition,
since the correction due to $M_x$ tends to improve the universality of 
the soft masses among different generations of squarks and sleptons, 
it does not make the situation for the FCNC problem worse.   

Finally, we should make a remark about an important effect on 
the soft scalar masses due to the hidden sector U(1) $D$-term.
As discussed in \cite{oneloop}, the hidden sector U(1) $D$-term 
can contribute to the soft scalar masses $m_{\tilde f}^2$ 
in the observable sector through the above discussed kinetic term mixing. 
Its contribution to $m_{\tilde f}^2$ is estimated as
\begin{equation}
(m_{\tilde f}^{D_h})^2=g_xg_hQ_fQ_\phi^{(h)}\tan\chi~
\langle\phi\rangle^2,
\end{equation}
where $g_h$ is the coupling constant of the hidden sector U(1) and 
$Q_\phi^{(h)}$ is its charge of the hidden sector field $\phi$.
As disdussed in the previous part, $\sin\chi$ should take values of 
$O(10^{-1})$ in the present scenario.
Thus, if $\langle\phi\rangle$ is much larger than $O(1)$~TeV, this
contribution dominates $m_{\tilde f}^2$.
In case of $Q_fQ_\phi^{(h)}<0$, in particular, it can make 
$m_{\tilde f}^2$ negative and upset the symmetry breaking
in the observable sector. This suggests that the present model requires 
the existence of rather light Abelian gauge field 
with the mass of $O(1)$~TeV in the hidden sector. 
This could be a typical feature of the model and
its effects might be examined in the future collider experiments. 
Thus, it seems  worth to make further study on this point 
and clarify its phenomenological effects in the observable sector. 
   
In summary we proposed the SUSY breaking scenario in which
different mediation effects of the SUSY breaking compete and 
the nonuniversal gaugino masses are induced in the Abelian sector. 
If there is the kinetic term mixing between the Abelian gauge fields in
the hidden and observable sectors, the Abelian gaugino in the observable
sector can have
additional contributions from this mixing in the framework of the 
ordinary gravity mediation SUSY breaking.
The main contribution of the SUSY breaking comes from the
gravity mediation except for the mass of the Abelian gaugino which 
has the kinetic term mixing with the Abelian gaugino 
in the hidden sector.
As an interesting phenomenological aspect of such a scenario, 
we studied the fine
tuning problem in the radiative symmetry breaking in the MSSM. We showed that
the RGE effects due to this gaugino mass can reduce the corrections to
the Higgs mass although the same effect increases the stop mass.
Since this SUSY breaking scenario can affect the neutralino 
phenomenology in the way that the lightest neutralino is dominated 
by the MSSM singlet fermion \cite{cdm}, the model may be examined 
through future collider experiments and dark matter searches.
  
\vspace*{5mm}

This work is partially supported by a Grant-in-Aid for Scientific
Research (C) from Japan Society for Promotion of Science (No.17540246).


\end{document}